\documentclass{ieeeaccess}
\usepackage{cite}
\usepackage{amsmath,amssymb,amsfonts}
\usepackage{algorithmic}
\usepackage{graphicx}
\usepackage{textcomp}
\def\BibTeX{{\rm B\kern-.05em{\sc i\kern-.025em b}\kern-.08em
    T\kern-.1667em\lower.7ex\hbox{E}\kern-.125emX}}
\begin{document}
\history{}
\doi{}

\title{Does a larger neural network mean greater information transmission efficiency?}
\author{\uppercase{Bartosz Paprocki}\authorrefmark{1}, 
\uppercase{Agnieszka Pregowska}\authorrefmark{2}, and \uppercase{Janusz Szczepanski}\authorrefmark{2}}

\address[1]{Institute of Mechanics and Applied Computer Science, Kazimierz Wielki
	University, Kopernika Str. 1, Bydgoszcz, 85–074, Poland (e-mail: bartekp@ukw.edu.pl)}
\address[2]{Institute of Fundamental Technological Research, Polish Academy of
	Sciences, Pawinskiego Str. 5B, Warsaw, 02–106, Poland (e-mail: aprego@ippt.pan.pl; jszczepa@ippt.pan.pl)}



\corresp{Corresponding author: Janusz Szczepanski (e-mail: jszczepa@ippt.pan.pl).}

\begin{abstract}
Realistic modeling of the brain involves a large number of neurons. The important question is how does this size affects transmission performance? Here this question is explored in terms of Shannon’s theory. Mutual Information between input and output signals for simple classes of networks with increasing number of neurons is analyzed theoretically and numerically. This type of analysis provides insight and intuition for more complex networks. It is clear that simple networks enable considering neural dynamics, while large complex networks are an efficient computational tool for statistical analysis. In this paper, the Levy-Baxter probabilistic neural model was applied. It turned out that for these simple networks the Mutual Information increases very slowly with the number of neurons. Moreover, our calculations show that for the practical number of neurons to which the input signal can reach at a given moment (up to 10,000), the values of Mutual Information reached only about half of the maximum value, which in our network is equal to the sum of the maximum information carried by single signals reaching the synapses of a given neuron. These results indicate that starting from a certain level (50 - 60 neurons), an increase in the number of neurons does not imply an essential increase in transmission efficiency, but it can contribute to reliability.

\end{abstract}



\maketitle

\section{Introduction}
\par
A human brain contains billions of neurons, linked to one another via hundreds of trillions of tiny contacts called synapses \cite{vanHemmenSejnowski2006}. It is known that more than 80\% of neurons are little branch cells located in the cerebellum and have received only a few electrical impulses (spikes) from 4-7 synapses, while the rest of the neurons have even up to 200 000 connections \cite{Shepherd1974}. Over the last two decades, huge progress has been made in explaining the evolution and the role of brain size \cite{Ascoli2022}. In the context of scale effects, analysis understanding how network size, specifically the number of neurons, influences information transmission efficiency is an important issue. 
\par
Neural activity at the microscopic level was modeled by phenomenological equations \cite{Gerstner2012}. Schwalger and co-authors proposed a system of equations for several interacting populations at the mesoscopic scale starting from a microscopic model of randomly connected generalized integrate-and-fire neuron models for networks varying between 50-2000 neurons \cite{Schwager2017}. In turn, in \cite{Barzon2022} was shown that structural networks are the crucial component in the stochastic brain model on the mesoscopic scale. The scale problems constitute also a challenge for efficient implementations and performance of advances networks. In \cite{Kunkel2012} linear models were considered to analyze the memory consumption of the constituent components of neuronal simulators as a function of network size and the number of cores used. The large-scale models of neuronal activity as describe whole neural populations' activity were considered in \cite{Ritter2022}. This kind of approach enables the integration of different information sources and analysis of the biophysiological mechanisms in the network. 
\par
Quantitative measuring of the information requires the application of adequate mathematical tools. In the Shannon Information Theory \cite{Shannon1948}, the neural networks are treated as communication channels and the information transmitted is measured as the Mutual Information between stimuli and response signals \cite{Shannon1948, Awan2019, Sengupta2022, Wen2022}.
\par
When studying the processing of information transmission, it is important to choose both neural models \cite{Gerstner2009} and network architecture models \cite{Park2013} first. In previous papers, we have studied directly transmission efficiency for simple neuronal ring architectures composed of a few Levy-Baxter neurons \cite{LevyBaxter2012} paying particular attention to the role of inhibitory neurons, long-range connections, and adaptation of neuronal networks to the presence of noise \cite{Paprocki20132,Pregowska2019}. This model of neurons has a probabilistic character and exploits the  binary representation of neuronal signals. Moreover, it contains all essential qualitative mechanisms participating in the transmission process and provides results consistent with physiologically observed values \cite{LevyBaxter2012}. 
\par
In this article, we focus on the problem of the influence of the number of neurons in the network on transmission efficiency. We analyze both theoretically and numerically the Mutual Information between input and output signals in the case of a simple class of neural networks with an increasing number of neurons. This type of analysis provides insight and intuition for more complex situations. We present results characterizing $MI$ dependence on the size of the network as well as on the adopted parameters of neurons. It is worth emphasizing that finding the maximum $MI$ actually means finding the so-called Shannon capacity of the transmission channel which directly characterizes optimal decoding opportunities. It turned out that for these simplified neural networks the maximum Mutual Information increases very slowly at the rate $m^{-c}$, with a small $c = 0.02473$, where $m$ is a number of neurons. This indicates among others that a further increase in the number of neurons does not imply a significant increase in transmission efficiency.

\section{Materials and Methods}
\par
It is commonly known that the carries of information are spike-trains. Taking into account the physiological issues connected with spike train appearance, each spike is detected with some limited time resolution. This led to the idea to represent spike trains by a sequence of symbols. The binary digitalization of spike trains is the most natural and commonly used representation \cite{Rieke1999}. Since a spike train is being observed with some limited time resolution $\delta$, so, in each time bin a spike is present (denoted by “1”) or absent (assigned by “0”) \cite{Rieke1999}. Then if we look at some time interval of length $T$, each spike train is represented by a binary sequence (additionally with some probability of occurence). Mathematically such sequences can be treated as a part of a trajectories of a stochastic process which can be analyzed with the use of Shannon Information Theory \cite{Pregowska2016}. The two fundamental concepts of this Theory are entropy and Mutual Information $(MI)$ between two random variables $\bf X$ and $\bf Z$ \cite{Shannon1948}. The concepts of entropy and Mutual Information have recently been intensively used in many problems related to the application of learning methods using neural networks in data classification problems \cite{Jo2022}. Mutual Information can be expressed in terms of entropies
\begin{equation} \label{Mutual_Information}
	MI(\bf X;Z):=H(\bf X)-H(\bf X|\bf Z)=H(\bf X)+H(\bf Z)-H(\bf X,Z),
\end{equation}
where $H(\bf X|Z)$ is the entropy of $\bf X$ conditional on $\bf Z$ and $H(\bf X,\bf Z)$ is the joint entropy of $\bf X$ and $\bf Z$ \cite{Bayram2012, Salafian2023}. Clearly, 0 $\leq MI(\bf X;Z) \leq H(X)$. 
\par 
The basic idea of Mutual Information is to determine the reduction of uncertainty (measured by entropy) of random variable $\bf X$ provided that we know the values of discrete random variables $\bf Z$. Maximal $MI$ is linked with the channel capacity, for a given communication channel through the Shannon Fundamental Theorem, which characterizes the optimal decoding schemes.

\Figure[t!](topskip=0pt, botskip=0pt, midskip=0pt)[width=1.0\linewidth]{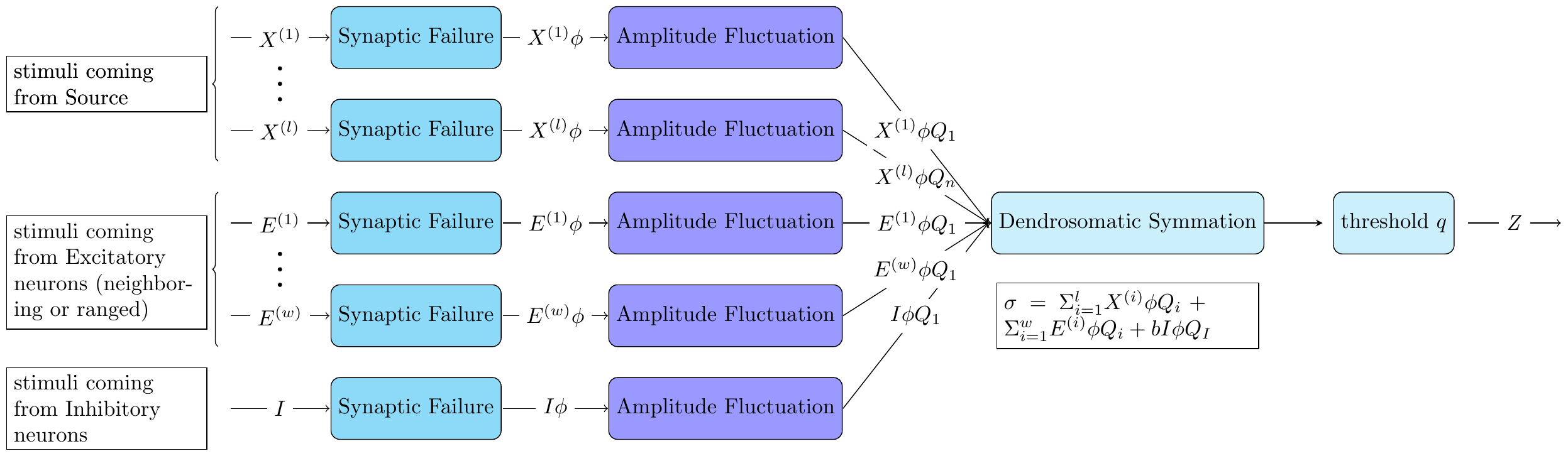}
{ \textbf{The scheme of the neuron model proposed by Levy and Baxter \cite{LevyBaxter2012}. Additionally, similarly to \cite{Paprocki20132} the type of inputs was also emphasized. Encoded stimuli is modeled by a discrete, binary stationary stochastic process with firing-rate $(f_{r})$ being the probability of a spike occuring and $1-fr$ the probability of no spike. Parameters $E^{(i)}$ describe excitatory strength, $b$ addresses inhibition strength, quantal failures $\phi$ is a random variable taking $0$ for the input $0$ and $1$ with probability $s$ for the input $1$, amplitude fluctuations $Q$ are implemented as random variables $U[0;1]$ with uniform distribution function. The activation threshold is denoted by $q$. When $\sigma$ is greater than $q$ a “1” bit is generated, otherwise a “0” is output.}\label{fig_neuron_model}}

\par
In this paper, we assume the binary representation of spikes and, a probabilistic Levy-Baxter neuron model \cite{LevyBaxter2012}, (see, Fig. \ref{fig_neuron_model}). In general, it takes into account all essential qualitative mechanisms, which are involved in the information transmission process. In this approach, the synaptic noise $s$ (success rate parameter), amplitude modulation $Q_{i}$ and activation threshold height $g$, are the neuron model's parameters. The binary input $\bf X=[\bf X^{(1)},…,X^{(n)}]$ to a given neuron and the binary output of the neuron are represented as a stochastic process, where $n$ denotes the number of synapses. To simplify the notation, we denote the inputs $E$ and $I$ by $\bf X$ with appropriate index. Here, as input stimuli, the Bernoulli stochastic processes were assumed. Thus, $H(\bf X)$ $\leq n$ and consequently $MI(\bf X;Z)$ must be less than number of synapses in a single neuron $n$. Each binary input is subject in synapses to quantal failures $\phi$ being Bernoulli distributed random variable (with parameter $s$) and quantal amplitude modulation $Q_{i}$ and all this is summed up to $\sigma$. This is the input to the spike generator $g(\sigma)$. The spike is generated if the magnitude of its excitation $\sigma$ exceeds the assumed threshold $g$. Thus, the output assumes as a binary stochastic process $\bf Z$.
\section{Results}
The purpose of this article is to give insight how the size of the network, i.e. the number of neurons, can affect transmission performance. To address this problem in terms of Shannon Theory we start by a brief recalling of the notation. Let $n$ denote the number of synapses for each neuron and $m$ the number of neurons in the network. The Information Source is assumed as a $\bf n$-dimensional stochastic process $\bf X(t_{i})=[X^{(1)}$$(t_{i}),…,\bf X^{(n)}$$(t_{i})],i=1,2,…$, and $t_{i+1}-t_{i}=\Delta$, where $\Delta$ is assumed time resolution. For example, one can assume that this information is coming from $n$ neurons and it is modeled by the processes $\bf X$. The components of the process $\bf X^{((i))}$ $(t_{i})$ are considered as independent Bernoulli processes with parameter $p$. This parameter can be understood as a firing rate $f_r$ of the input spike trains to a single synapse of a neuron located in the second layer $\bf N$. The joint neuronal output from the layer $\bf N$ of $m$ neurons (which is a $m$-dimensional random variable) was denoted by $\bf Z=[Z^{(1)},…,Z^{(m)}]$, where $\bf Z^{(i)}$ is a binary random variable (see, Figure \ref{fig_network_model}).

\Figure[t!](topskip=0pt, botskip=0pt, midskip=0pt){ModelNeuronuSiec}
{ \textbf{The architecture of the neural network under consideration. Each of neurons $X_{i}$, $i=1,2,…,n$ in the input layer $\bf X$ is supported by signals coming from some other sources, e.g. from earlier neurons. Each neuron $N_{j},j=1,2, …,m$ in the second layer $\bf N$ is supported by inputs coming from Information Source  $\bf X$ (constituting by $n$ neurons). Thus, here we assume that the neurons in the second layer have $n$ synapses. In the paper the Mutual Information $ MI \bf(X,Z)$ between the information delivered by neurons from layer $\bf X$ and information carried out by the output layer $\bf Z$ is evaluated and analyzed.}\label{fig_network_model}}

\subsection{The theoretical analysis}
\par
Now, let’s assume that $\bf x^{k}$ is the event that at a given moment of time, $k$ of specific components $\bf X^{(i)}$ in $\bf X$  being inputs to neurons in layer $\bf N$ are equal to $1$, and the other $n-k$ is equal to $0$. Similarly, let $\bf z^{j}$ is the event that at a given moment of time, $j$ of specific components $\bf Z^{(i)}$  in $\bf Z$ being output from the layer $\bf Z$ are equal to $1$, and the other $m-j$ are equal to $0$. 
\par
Since the random variables $\bf X^{(i)}$, are independent, thus the probability of the event $\bf x^{k}$ reads
\begin{equation} \label{event_probability}
	P(\mathbf{x^{k}})=P(\mathbf{X=x^{k}}) = f_{r}^{k}(1-f_{r})^{n-k}.
\end{equation}
For each neuron from the output layer, the input spike can pass through the synapse with the success rate $s$. Next, the amplitudes of the transmitted signals are modulated by random function $Q$ with the uniform distributions on the interval $[0;1]$. Thus, the conditional probability of activation of a single neuron provided that the event $\mathbf{x^{k}}$ occurs is
\begin{equation} \label{activation_single_neuron_1}
	P(\mathbf{Z=z^{1}|X=x^{k}})= \\ \sum_{i=0}^{k} {{k}\choose{i}} s^{i}(1-s)^{k-i}P(iQ \geq g)
\end{equation}
\begin{equation}\label{activation_single_neuron_0}
	P(\mathbf{Z=z^{0}|X=x^{k}})= 1 - P(\mathbf{Z=z^{1}|x^{k}})
\end{equation}
where $P(iQ \geq g)$ denotes the probability that the sum $iQ$ of $i$ random variables of the type $Q$ reaches the activation threshold $g$ and $z^{1}$ is the probability of activation of a single neuron. Since each $Q$ is uniformly distributed and independent, thus the random variable $iQ$ has the Irwin-Hall distribution \cite{Hall1927}. Then, the probability $P(iQ \geq g)$ can be given as 
\begin{equation}\label{irwinhall1}
	P(iQ \geq g)=1-P(iQ < g),
\end{equation}
where the cumulative distribution function (CDF) is of the form \cite{Hall1927}
\begin{equation}\label{irwinhall2}
	P(iQ < g)= \frac{1}{i!}\sum_{h=0}^{|g|} (-1)^{h}{{i}\choose{h}}(g-h)^{i}.
\end{equation}
Note, that calculation of the probability (\ref{irwinhall1}) is directly obtained from the CDF of $iQ$ at point $g$. By substituting formulas (\ref{irwinhall1}) and (\ref{irwinhall2}) to (\ref{activation_single_neuron_1}) we have
\begin{eqnarray} \label{component1}
	P(\mathbf{Z=z^{1}|X=x^{k}})= \\ \nonumber 
	\sum_{i=0}^{k} {k\choose i} s^{i}(1-s)^{k-i}(1-\frac{1}{i!}\sum_{h=0}^{|g|} (-1)^{h}{{i}\choose{h}}(g-h)^{i}).
\end{eqnarray} 
Since the components in the output $\mathbf{Z}=[\bf Z^{(1)},…,Z^{(m)}]$ are independent, thus we have 
\begin{eqnarray}\label{component2}
	P(\mathbf{Z=z^{j}|X=x^{k}})= \\ \nonumber (P(\mathbf{Z=z^{1}|X=x^{k}}))^{j} (	P(\mathbf{Z=z^{0}|X=x^{k}}))^{m-j},
\end{eqnarray}
\begin{equation} \label{component3}
	P(\mathbf{Z=z^{j},X=x^{k}})=P(\mathbf{x^{k}}) P(\mathbf{Z=z^{j}|X=x^{k}})
\end{equation}
\begin{equation} \label{component4}
	P(\mathbf{Z=z^{j}})=\sum_{x \in X}P(\mathbf{x,z^{j}})= \\  \sum_{k=0}^{n}{{n}\choose{k}}P(\mathbf{X=x^{k},Z=z^{j}}).
\end{equation}
Thus, we can calculate all components (i.e. corresponding entropies) which are needed to determine the Mutual Information $MI(\mathbf{X;Z})$ in (\ref{Mutual_Information}). These entropies are expressed as 
\begin{eqnarray}
	H(\mathbf{X})=H(\mathbf{X}^{(1)})+H(\mathbf{X}^{(2)})+...+H(\mathbf{X}^{(n)})=\\ \nonumber -n [f_{r} \log f_{r}+(1-f_{r})\log (1-f_{r})],
\end{eqnarray}
\begin{eqnarray}
	H(\mathbf{Z})=-\sum_{z \in Z}P(\mathbf{z})\log P(\mathbf{z})= \\ \nonumber - \sum_{j=0}^{m}{{m}\choose{j}}P(\mathbf{z^{j}})\log P(\mathbf{z^{j}}),
\end{eqnarray}
\begin{eqnarray}
	H(\mathbf{Z|X})= \\ \nonumber -\sum_{x \in X}\mathbf{\sum_{z \in Z}} P(\mathbf{z,x})\log P(\mathbf{z|x})= \\ \nonumber - \sum_{k=0}^{n}{{n}\choose{k}}P(\mathbf{x^{k}})\sum_{j=0}^{m}{{m}\choose{j}}P(\mathbf{z^{j}|x^{k}})\log P(\mathbf{z^{j}|x^{k}}),
\end{eqnarray}
\begin{eqnarray}
	H(\mathbf{X,Z})=-\sum_{x \in X}\mathbf{\sum_{z \in Z}} P(\mathbf{x,z})\log P(\mathbf{x,z})= \\ \nonumber - \sum_{k=0}^{n}\sum_{j=0}^{m}{{n}\choose{k}}{{m}\choose{j}}P(\mathbf{x^{k},z^{j}}) \log P(\mathbf{x^{k},z^{j}}) .
\end{eqnarray}
Thus, the Mutual Information $MI\mathbf{(X;Z)}$ is of the form 
\begin{eqnarray}\label{MI}
	MI\mathbf{(X;Z)}=-n [f_{r} \log f_{r}+(1-f_{r})\log(1-f_{r})] + \\ \nonumber-  \sum_{j=0}^{m}{{m}\choose{j}}P(\mathbf{z^{j}})\log P(\mathbf{z^{j}})+ \\ \nonumber + \sum_{k=0}^{n}\sum_{j=0}^{m}{{n}\choose{k}}{{m}\choose{j}}P(\mathbf{x^{k},z^{j}}) \log P(\mathbf{x^{k},z^{j}})
\end{eqnarray}
and it can be calculated by substituting (\ref{component2}), (\ref{component3}), and (\ref{component4}) into (\ref{MI}).
\par
In the next Section, we apply formula (\ref{MI}) to find the Mutual Information between the input signals $\mathbf{X}$ and output signals $\mathbf{Z}$ for the considered simple networks with the increasing number of neurons m, perform calculations for the full range of parameters characterizing a Levy-Baxter neuron.
\subsection{Numerical simulations}
We have done a numerical simulation to evaluate $MI(\mathbf{X;Z})$ by exploiting the formulas (\ref{component2}), (\ref{component3}), (\ref{component4}), and (\ref{MI}) developed in the previous section. To find maximal $MI$ with a satisfactory accuracy we needed to go through the neuron parameter space of synaptic failure $s$ being $0<s<1$ and through firing frequency $f_{r}$ $(0<f_{r}<1)$ with a relatively small step equal to 0,01.  Since the Levy-Baxter model of neuron has a probabilistic nature, simulating the input-output process for a single neuron requires the use of randomizing generators (working according a given probability distribution). This fact means that $MI$ calculations can be successfully performed for a network containing at most up to several dozen of neurons. To go during the computations with the number of neurons $m$ possibly large and taking into account the information concerning number of synapses in \cite{Shepherd1974} we performed the analysis for neurons with 5 synapses. The threshold parameter $g$ is assumed to be 5\% of the maximal possible value that can be reached by the neuron with 5 synapses.  Despite these limitations, our results showed quantitative and qualitative behavior of maximal $MI(\mathbf{X;Z})$ as a function of the number of neurons ($m$). The results are presented in Figure 3, Figure \ref{izo}, and Table \ref{tab1}.

\Figure[t!](topskip=0pt, botskip=0pt, midskip=0pt)[width=1.0\linewidth]{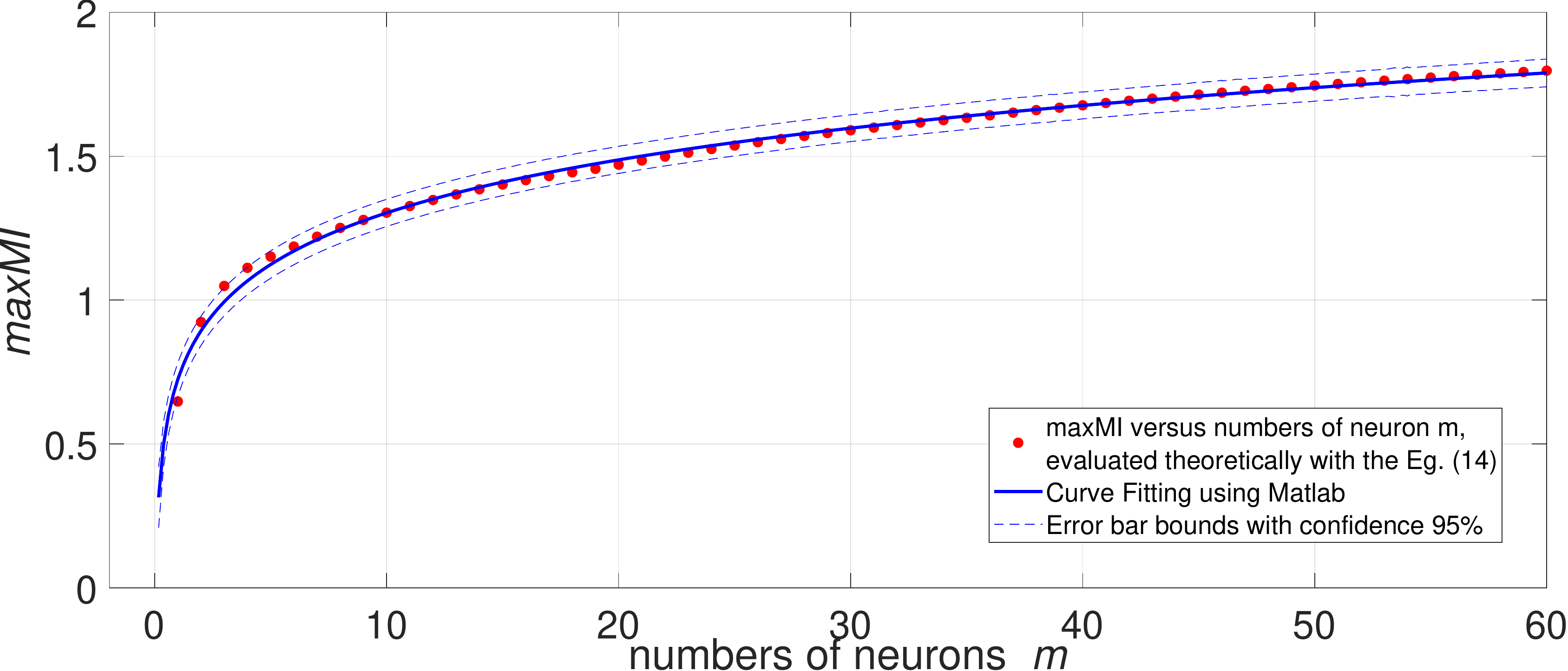}
{ \textbf{The influence of an increasing number ($m$) of neurons in the layer $N$  on maximal Mutual Information $MI\mathbf{(X;Z)}$. For each number m the maximum was taken over the source (i.e. layer $X$) parameter $f_{r}$ and neuron parameter $s$ being the success rate (i.e. probability $s$ that spike will be successfully transmitted over a given synapse) in layer $N$. Activation threshold $g$ is assumed to be 0.25. The calculations are presented for $n$=5, what means that the number of synapses of neurons in layer $N$ is already 5. It turned out that the best fitting curve is of the form $f(m)=a-\frac{b}{m^{c}}$ $(a=12.05, b=11.37, c=0,025)$, where $m$ is the number of neurons in the Layer $N$. The goodness of fitting is: $SSE:0.02096, RMSE: 0.01917$. The confidence bounds with 95\% of confidence level are also depicted ($\pm 2SD$).}\label{fig_maxMI}}

\par
In Figure 3 the influence of the increasing number of neurons $m$ on maximal Mutual Information $MI(\mathbf{X;Z})$ for neural networks with architecture presented in Figure 2 is shown. Applying MATLAB Curve Fitting Toolbox we found that the best fitting curve is the function $maxMI(m)=12.05-\frac{11.37}{m^{0.025}}$  with a goodness of fit $RMSE=0.019$ and $SSE=0.02$. This shows that $maxMI$ is asymptotically limited and another important observation is that it increases very slowly starting from the number of $m$ about 50 – 60. Moreover, since we assumed in our simulations that the number of synapses for each neuron is equal to 5 and that the input signals to each synapse come from the Bernoulli process, taking into account the classical inequality $MI(X;Z)\leq H(X)$, we have that $MI( X;Z)$ can be up to 5. So we see that for a practical number of neurons $m$ up to 10000, the value of $MI(X;Z)=f(m) \sim 2.57$ is only about half of the maximum possible value. 
\par
In Figure \ref{izo} Mutual Information $MI\mathbf{(X;Z)}=MI(f_{r},s,g)$ as a function of source parameter firing rate $f_{r}$ and neuron parameter synaptic noise $s$ for increasing number $m$ (5, 10, 30, 60) of neurons in layer $\mathbf{N}$ is presented. One can observe that with the increase of the network size $m$ the values of $maxMI$ (in another words the capacity of the transmission channel) are reached for smaller $s$ and larger $f_{r}$ (see also Table \ref{tab1}). This observation confirms that more energy needs to be put on to obtain still more efficient transmission with increasing size and as some recompensation this maximum is reached even for more noisy channel (i.e. for lower success rate $s$).

\Figure[t!](topskip=0pt, botskip=0pt, midskip=0pt)[width=1.0\linewidth]{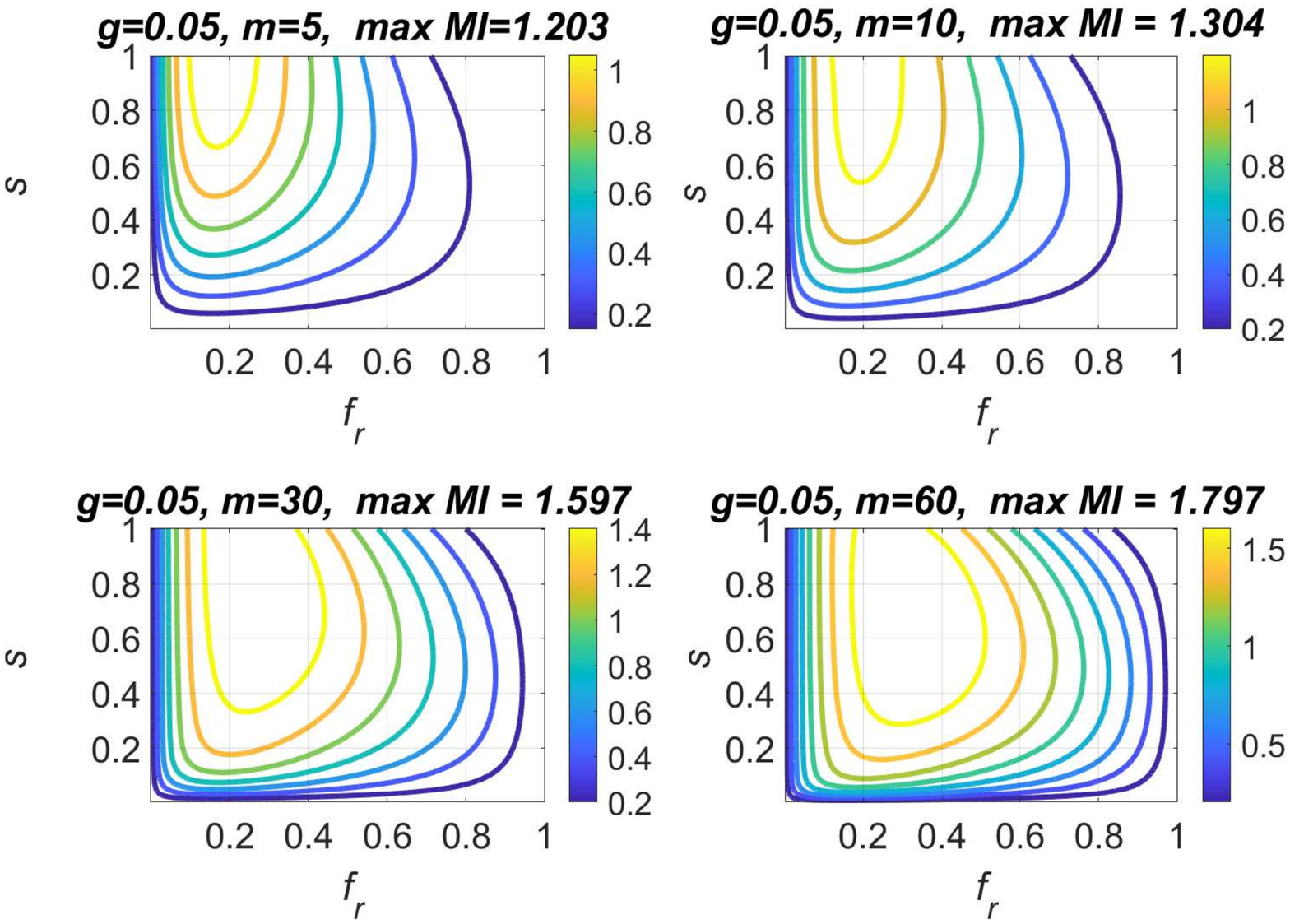}
{ \textbf{Mutual Information $MI\mathbf{(X;Z)}=MI(f_{r},s,g)$ for neural networks shown in Figure 2 stimulated by a Bernoulli information source with firing rate $f_{r}$ (horizontal axe) and synaptic noise $s$ (vertical axe). The number $m$ of neurons in the second layer $N$ increases successively, $m=5,10,30,60$. The isolines with step 0.05 are depicted.}\label{izo}}

\begin{table}
	\begin{center}
		\begin{tabular}{ |c|c|c|c|c| } 
			\hline
			network size $m$ & threshold $g$ & $maxMI$ & $s$ & $f_{r}$ \\ 
			\hline 	
			5 & 0.050 & 1.203 & 0.980 & 0.172 \\ 
			\hline
			10 & 0.050 & 1.304 & 0.900 & 0.199 \\ 
			\hline
			30 & 0.050 & 1.597 & 0.821 & 0.262 \\ 
			\hline
			60 & 0.050 & 1.797 & 0.695 & 0.317 \\
			\hline
		\end{tabular}
	\end{center}
	\caption{Maximal Mutual Information for selected number of neurons $m$ (5, 10, 30, 60). The parameters $s$ and $f_{r}$ for which these maxima are achieved are also given.} \label{tab1}
\end{table}
\section{Disscusion and Conclusions}
\par
In the course of the evolution process mechanisms have been developed to enable more and more efficient and reliable information processing. On the other hand, the key question is the impact of the size of the brain and thus the role of the number of neurons on the efficiency of these processes. Can such performance be significantly improved by a simple increasing the size of the neural network? To provide an insight into this problem, it seems natural to analyze these issue based on examining the relevant models of networks and neurons themselves \cite{Tiesinga2001, Kitazono2020}. Still, the realistic models of real neural networks are analytically and computationally intractable. One of the major difficulties is the selection of the proper size and topology of these networks. Hunter and co-authors \cite{Hunter2012} discussed hot issue, including different learning algorithms, the efficiency of different network topologies, and the importance of choosing the proper size of neural networks. While in \cite{Buice2013} a new formalism that borrows from the methods of many-body statistical physics to analyze finite-size effects in spiking neural networks was introduced. 
\par
It is known that traditional mathematical approaches to analytically studying the dynamics of neural networks rely on mean-field approximation, which is rigorously applicable only to infinite-sized networks \cite{Fasoli2018}. However, all existing real biological networks consist of a finite number of neurons, often consisting of only a few dozen neurons, such as microscopic circuits in invertebrates. Therefore, it is important to be able to extend our ability to analytically study neural dynamics to small networks. At present, systematic analytical solutions to the dynamics of neural networks of finite sizes still require further analysis. 
\par
In this paper, in order to give insight into the problem of the influence of network size on information transmission efficiency, we consider a simple networks consisting of Levy-Baxter neurons. We treat this problem by using the Shannon approach and analyzing the Mutual Information between input and output signals. The L-B neuron, which exhibits the basic properties of a biological neuron, is described in probabilistic language, which allowed us to find analytical formulas for $MI$ expressed in terms of the size of the network and the parameters of the neuron. 

Numerical simulations using these formulas have shown that for practical number of neurons to which the input signal can reach at a given moment (up to 10,000), the Mutual Information between input and output signals is about 50\% of the maximum possible information that can be achieved, and it is also important that the increase in $MI$ is very slow as the neurons number increases. This suggests that large number of neurons in actual biological networks (brain) is related mostly to the fact that individual areas in the brain are dedicated to different types of stimuli and it is rather due to the tendency to achieve reliability and noise immunity rather than a significant increase in the information performance.
\appendices
\EOD

\end{document}